\documentclass[aps,superscriptaddress,altaffilletter,tightenlines]{revtex4}

\begin{document}


\title{Mathematical Features of Several Cosmological Models}

\author{Luis P. Chimento and Alejandro S. Jakubi                \\
{\it Departamento de F\'{\i}sica, Facultad de Ciencias Exactas y Naturales, }\\
{\it Universidad de Buenos Aires, Ciudad  Universitaria,  Pabell\'{o}n  I, }\\
{\it 1428 Buenos Aires, Argentina.}}

\begin{abstract}
Einstein equations for several matter sources in homogeneous, isotropic metric are shown to reduce to a second order nonlinear ordinary differential equation. An analysis of its solutions is made in an important case. 
\end{abstract}
\maketitle
 

\section{Introduction}

Exact solutions of the Einstein equations are difficult to obtain due to their
nonlinear nature. In this paper we show that the system of
equations for homogeneous, isotropic cosmological models with a
variety of matter sources reduce to particular cases of the ordinary
differential equation

\begin{equation} \label{1}
\ddot y+y^n\dot y+\beta y^{2n+1}+cy^n =0
\end{equation}

\noindent
where $\beta$, $c$ and $n$ are constants.

The problem of a causal viscous fluid with the bulk viscosity coefficient
$\zeta$ proportional to $\rho^{1/2}$ corresponds to $n=1$, $c=0$, $y\propto H$
in the truncated Extended Irreversible Thermodynamics theory,\cite{visco} and
$n=-1/r$, $c=0$, $y\propto H^{-r}$ in the full theory, where the relation
between temperature and energy density is assumed to be of the form $T\sim
\rho^r$.\cite{Maar} Also, the behavior near the singularity, when the
relaxation term is much more important than the viscous term in the transport
equation, corresponds to $n=1$, $c=0$, $y\propto H$ for generic power-law
relation $\zeta=\alpha\rho^m$.\cite{Zak} For a perfect fluid source, with an
equation of state $p=(\gamma-1)\rho$, and a cosmological constant $\Lambda$,
we recover Eq.(\ref{1}) with $n=1$, $c=-\gamma\Lambda$ provided we derive
twice the $00$ Einstein equation

\begin{equation} \label{2}
H^2=\frac{1}{3}\frac{\rho_0}{a^{3\gamma}}-\frac{k}{a^2}+\frac{\Lambda}{3}
\end{equation}

\noindent We also find Eq.(\ref{1}) with $n=1$, $c\propto\Lambda$, $y\propto
H$ for a fluid with equation of state $3p=-\rho-C/a^2$, $\rho=C \log a/a^2$,
$C$ a constant. For two scalar fields, one free and the other
self--interacting with a potential $V(\phi)=V_0\exp(-A\phi)$, the change
$y=a^{A^2/2}$ in the Einstein equation leads to Eq.(\ref{1}) with $n=-6/A^2$
and $c=0$, as can be seen in the paper "Two-scalar field cosmologies"
appearing in this volume. For a time decaying cosmological "constant",
$\dot\Lambda\sim -H^3\Lambda$ with $y\propto H$, the case with $n=1$, $c=0$
also arises.\cite{Reut}

Thus, it turns out to be of great interest to analyze Eq.(\ref{1}) from the
mathematical point of view. Its general solution will be studied elsewhere, and
we concentrate here on the families of real solutions of the case $n=1$, $c=0$.

\section{Analysis of the solutions for $n=1$, $c=0$}

Unless $\beta=1/9$, Eq.(\ref{1}), for $n=1$ and $c=0$ has only two point Lie
symmetries and it is not equivalent to a second  order  linear  equation
under  a point  transformation.\cite{Lea}  So,   we   consider   the
nonlocal transformation

\begin{equation} \label{21}
z = y^2 ,\qquad \eta =\int   y dt
\end{equation}

\noindent
which turns Eq.(\ref{1}) into the equation of a damped linear oscillator

\begin{equation} \label{22}
\frac{d^2z}{d\eta^2} + \frac{dz}{d\eta} + 2\beta z = 0
\end{equation}

\noindent
and we obtain the general solution  of  in  a  parametrized
form $(t(\eta),y(\eta))$. The real solutions of (\ref{22}) can be classified as
follows:

\noindent
a. $\beta<1/8$ (strong damping).
\begin{equation} \label{23}
z(\eta) = C \exp (\lambda_{\pm}  \eta)
\end{equation}
\begin{equation} \label{24}
z(\eta) = 2C \exp (- \eta/2) \cosh \left(\delta \eta/2 + \phi\right)
\end{equation}
\begin{equation} \label{25}
z(\eta) = 2C \exp (- \eta/2) \sinh \left(\delta \eta/2 + \phi\right)
\end{equation}
\noindent
b. $\beta=1/8$ (critical damping).
\begin{equation} \label{26}
z(\eta) = C \exp (- \eta/2)
\end{equation}
\begin{equation} \label{27}
z(\eta) = C (\eta + \phi) \exp (- \eta/2)
\end{equation}
\noindent
c. $\beta>1/8$ (weak damping)
\begin{equation} \label{28}
z(\eta) = 2C \exp (- \eta/2) \sin (\delta \eta/2 + \phi)
\end{equation}
\noindent
where $\lambda_{\pm}$   are  the  roots  of  the  characteristic  polynomial,
$\delta=|1-8\beta|^{1/2}$, and $C$, $\phi$ are arbitrary integration
constants.
Through the transformation (\ref{21}),  both  Lie  point  symmetries  of
(\ref{1}) have a simple equivalent: $t(\eta)$ is defined up to an  arbitrary
integration constant $t_0$ , and this reflects the invariance of  (\ref{1})
under $t\to t+t_0$ . Also,  the  invariance  $z\to z/A^2$   $(A\neq 0)$,  of
(\ref{22})  is
equivalent to the symmetry transformation $t\to |A|t$, $y\to y/|A|$. Besides,
the permutation between the two  branches  of  $z$     leads  to  the
discrete symmetry transformation $y\to -y$, $t\to -t$.

Whenever $z(\eta)$ has a zero, extremum or inflexion point at  $\eta_1$ ,
$y(t)$
has a zero, extremum or inflexion  point  at $t_1 =t(\eta_1 )$. Besides,  it
can be seen that $\dot y$ is finite at any zero point, and so $y(t)$ is odd  in  a
neighborhood of all finite zero points (see below).

There are two groups of solutions $z(\eta)$:

\noindent
i) Those that never vanish, i.e., (\ref{23}),(\ref{24}), and (\ref{26}),  so
that we may choose $z(\eta)>0$ for all $\eta$. For these  solutions  $y(t)$  is
nonvanishing, and it is obtained from any of the  two  branches  of 
$\sqrt{z}$ (depending on sign $y$).

\noindent
ii) Those that have (at least) one  zero point,  i.e.,  (\ref{25}),
(\ref{27}), and (\ref{28}). The requirement that $z(\eta)>0$ cannot be
satisfied on  both sides of the zero point by the same solution (with a given value
of $C$). Therefore, $z(\eta)$ gives rise to two solutions y(t), one for each
sign of $C$. Since $y(t)$ is odd (see bellow), these solutions are obtained by
joining at the zero point both branches of $\sqrt{z}$.

From (\ref{21}), we see that $\eta(t)$ is even for odd $y(t)$ and  has extrema
at the zero points of $y(t)$. Then, for the group (i), $\eta(t)$ is monotonic, that
is, there is a one to one mapping between the real axis $\eta$ and some
interval of the axis $t$. However , for the group (ii), the pair of branches
at each  side  of  the  zero point correspond  to  different mappings between
$\eta$ and $t$. Besides, each singularity of $y(t)$  (where $\eta(t)$ diverges
logarithmically) marks a boundary for  the mapping $\eta\to t$. For solutions
(\ref{24}), (\ref{25}), (\ref{26}), and (\ref{28}), $t(\eta)$  can be
expressed in
terms of a hypergeometric function. Only for $\beta=1/9, 0$ or $-1$,
$t(\eta)$ can be inverted in closed form.

Due to the symmetries of (\ref{1}), if $y(t)$ is a  solution,  $A y(A\Delta
t)$  is 
also a solution, where $\Delta t=t-t_0$ . In particular,  if  $y(\Delta t)$  satisfies
$y=0$ and $\dot y\neq 0$ at $t=t_0$ , $-y(-\Delta t)$  is  also  a  solution  with  the  same
initial data. However, as (\ref{1})  satisfies  a  Lipschitz  condition,
given these initial data the solution is unique. Thus,  we  conclude 
that $y(\Delta t)$ is odd. Further, it is easy to see that $y(t)$ is  analytic
at $t_0$ , so  that  there  is  an  interval  where  its  Taylor  series
converges. As $y(t)$ must be odd also about any  further  zero point within
the interval of convergence of the series, it comes out  that  there 
are only two possibilities for an interval with a zero point:

\noindent
a) The interval contains only one zero point.

\noindent
b) There are infinitely many equispaced zero points; that is, $y(t)$ is an
oscillatory  periodic  function  (the  radius  of  convergence  is 
infinite). 

Solutions which exhibit behavior (a) occur only for $\beta<1/8$,  while  the  solutions  for
$\beta>1/8$ have behavior (b). So, for $\beta<1/8$,
$y(t)$ 
has either one or no zero point in the interval where it is bounded.

The  solutions  (\ref{23})  lead  to  the  two  one-parameter  families
of solutions for $\beta<1/8$:
\begin{equation} \label{31}
y_{\pm} (t) = \alpha /\Delta t  ,\quad  \alpha_{\pm} = -2/\lambda_{\pm},\quad
\lambda_+\neq 0;\qquad
y_+  = K ,\quad  \lambda_+  = 0
\end{equation}
\noindent
and we wish to investigate small departures from them.   Let   us   consider   first   the   case   when
$\exp[(\lambda_- -\lambda_+ )\eta-2\phi]\ll 1$. As $\lambda_+ >\lambda_-$ ,
this occurs for any $\phi$  if  $\eta$  is  big
enough. Then to first order we get the approximated solution
\begin{equation} \label{33}
y(t) = \frac{\alpha}{\Delta t} \left( 1 + \gamma {\Delta t}^r  \right),\quad
\lambda_+\neq 0;
\qquad
y(t) = K \left(1 + \gamma \exp(-K t) \right   )  ,\quad \lambda_+  = 0
\end{equation}
\noindent
where $\gamma\propto\exp(-2\phi)$, $r=4-\alpha$ is the
Kowalevski exponent and $\alpha=\alpha_+$  in this case.\cite{Yosh} In the opposite
case, that is for $\eta\to -\infty$,  we get also (\ref{33}a), but now
$\alpha=\alpha_-$  and $\gamma\propto \exp(2\phi)$. Whenever
$y(t)$ has a singularity, $z(\eta)$ diverges, and this occurs for
$\eta\to\infty$ as well as for $\eta\to -\infty$ if $\beta<0$. Hence, using
(\ref{21}),  we  find  that any singularity is located at a finite time, and
in a  neighborhood of it, $y(t)$ has the asymptotic form (\ref{33}a) with
$r>0$. When $\beta>0$, $z(\eta)\to 0$ and $|t(\eta)|\to\infty$  for
$\eta\to\infty$. Then,  $y(t)$  vanishes  at infinity with the asymptotic
behavior (\ref{33}a), where $\alpha=\alpha_+$   and  $r<0$.

The two-parameter families of solutions arise  from  (\ref{24}),  (\ref{25}),
(\ref{27}), and (\ref{28}). We classify  them in two groups: those which have
a singularity at a finite time, and those which are regular for all 
time; and we give the main features of their behavior.

\noindent
a. Singular solutions.

\noindent
$0\le\beta<1/8, C>0$.

\noindent
a1. They have a singularity, where the leading  behavior is (\ref{31}--),
decrease monotonically and vanish at infinity with leading behavior 
(\ref{31}+) unless $\beta=0$, when they have a nonvanishing limit.

\noindent
$\beta<0, C>0$.

\noindent
a2. They have two singularities, one  with  leading
behavior (\ref{31}--) and the other with leading behavior
(\ref{31}+). There is a minimum (maximum) between them.

\noindent
a3. They have a zero point between two singularities, where the leading
behavior is (\ref{31}+). They  increase  monotonically  and  have
three inflexion points.

\noindent
$\beta\le 1/8, C<0$.

\noindent
a4. They have a zero point between two singularities, where the
leading behavior is (\ref{31}--) for $\beta<1/8$ and
\begin{equation} \label{39}
y(t)\sim \frac{4}{\Delta t} \left[1+\frac{1}{2\ln |\Delta t|}+
\frac{A-\ln\ln(1/|\Delta t|)}{\left(2\ln|\Delta t|\right)^2}\right]
\end{equation}
for $\beta=1/8$.  They   decrease   monotonically.

\noindent
b. Regular solutions.

\noindent
$0<\beta\le 1/8, C>0$.

\noindent b1. They have a zero point between two extrema. They have  three
inflexion  points and they vanish at infinity, with leading behavior
(\ref{31}+)  for  $\beta<1/8$  and  like a4., with   the   replacement
$\ln(1/|\Delta t|)\to \ln|\Delta t |$, for $\beta=1/8$.

\noindent
$\beta=0, C>0$.

\noindent
b2. They have a zero  point at  $t_0$   and  increase  monotonically  with  a
nonvanishing limit at infinity.

\noindent
$\beta>1/8$.

\noindent
b3.
They are oscillatory periodic, and its period and amplitude have a
relation of the form $AT^2=f(\delta)$.
The  period  diverges as $\beta\to 1/8^+$  and has the limit $T\to
2\pi^{3/2}/(\sqrt{C}|\Gamma(3/4)|^2)$, $\beta\to\infty$.
These results agree with the phase space analysis,\cite{Min} and confirm
the numerical simulations.\cite{Lea2}

\section{ Conclusions}

We have obtained the general solution of Eq.(\ref{1}) for $n=1$ and $c=0$ in
a parametrized
form by means of Eq.(\ref{21})
The solutions  have  moving  singularities  and
depending on whether these points are real or not, two
groups of real solutions arise: the singular and the regular ones.\cite{Ince}
Both one-parameter families of  solutions for $\beta <1/8$  are
singular, unless $\beta =0,$ when one of them turns into a constant. They
coalesce for $\beta =1/8,$ and there is no real one-parameter  family  of
solutions  for $\beta >1/8.$  These  one-parameter  solutions  give  the
leading  behavior  of  the   solutions   about   a
singularity. Only for $\beta=1/9,0,-1$ two-parameter solutions are functions
on the complex plane and real solutions can be expressed in closed form.

In general, the problem of the construction of explicit  solutions
of a given  integrable  nonlinear  differential  equation  remains
open. One direction along which one  can  attempt  to  proceed  is
linearization, i.e. the reduction of  the  equation  to  a  linear
ordinary  differential  equation,   which   is,   by   definition,
integrable.
Only for $\beta =1/9$ Eq.(\ref{1}) possesses eight  symmetries  and
is linearizable by a point transformation. On  the  other
hand, the transformation (\ref{21}) linearizes it  for  any
value of $\beta$. Thus, although it has only two Lie  point  symmetries,
it possesses eight nonlocal symmetries. We think  that  it  is  of
utmost   importance   to   study   this   kind   of    linearizing
transformations, which have received up to  now  little  attention.

\end{document}